\documentstyle[12pt]{article}
\newcommand{\xq}{\begin{equation}}
\newcommand{\zq}{\end{equation}}
\newcommand{\beq}{\begin{eqnarray}}
\newcommand{\eeq}{\end{eqnarray}}
\title{{\hfill\normalsize ITP-95-21E}\\[1.0cm]
Dimensional Reduction in Nambu--Jona--\\Lasinio Model in
External Chromomagnetic Field}

\author{{\sl I.A. Shovkovy and V.M.Turkowski}\\
{\sl Bogolyubov Institute for Theoretical Physics,
252143 Kiev, Ukraine}}
\date{July 1, 1995}
\begin{document}
\maketitle

\vfill

\begin{abstract}
The effective dimensional reduction of fermion-antifermion
pairing dinamics in Nambu-Jona-Lasinio model in external
chromomagnetic field is considered. It is shown that such a
field causes the reduction of the space dimension by one
unit in contrast to an ordinary magnetic field which reduces
it by two units. The enhancement of the dynamical chiral
symmetry breaking due to the reduction is discussed.
\end{abstract}
\vfill
\eject

Relativistic models of massless fermions with four-fermion (4F) interaction
have become very popular mainly due to their applicability to the low energy
dynamics of pions. The widespread use of these models is also accounted for
the relative simplicity of them for theoretical studying. The second argument
was also decisive in our choosing the NJL model. The reduction which we
found in NJL model should take place in any fermionic model since the reason
underlying such a reduction has a general nature, though the particular
its manifestation may be different.

The characteristic feature of models with 4F interaction is the appearence
of the dynamical chiral symmetry breaking (D$\chi$SB) at 4F coupling
exceeding some critical value. It has
been just this property which has been utilized in the original paper
by Nambu and Jona-Lasinio (NJL). Since that time a great deal of other
papers devoted to studying models with 4F interaction have appeared. Some
of the references are \cite{GrNv,RosWarPar,Volkov}. In addition to them,
recently, interest to 4F models in different external fields
considerably arose \cite{Klev,Vshiv,GMS1,GMS2}. Such an interest is
mainly accounted for two reasons: 1) these models persist to be good
toy examples for modeling some more difficult problems ( such as QCD
vacuum \cite{Vshiv}, confinement and other) 2) they could also be
very useful for investigating the planar phenomena ( such as
antiferromagnetism \cite{And}
or superconductivity \cite{RosWarPar} ).

In the present paper we study the phenomenon of the effective dimensional
reduction of the fermion-antifermion pairing dynamics in the external
magnetic and chromomagnetic fields. A similar phenomenon was found for
fluctuations in superconductors near the critical line by Lee and Shenoy
long ago \cite{LeeSh}. Recently, the idea of dimensional reduction has been
fruitfully applied to the phenomenon of D$\chi$SB in several relativistic
models \cite{GMS1,GMS2,GMS3}. In those papers the dimensional reduction
manifested itself through the enhancement of D$\chi$SB when an external
magnetic field was applied to the system. The explaination of such an
enhancement is very simple. It is based on the well known fact that
low dimensionality promotes D$\chi$SB in the leading order in $1/N$
perturbation theory. However, when the number of dimensions becomes
less than $1+1$ ( or $2+1$ in the case of continuous symmetry ) the
possibility of symmetry breaking disappears due to the fameous
Coleman theorem ( to see that one need to take into account next to
leading order in $1/N$ perturbation theory ). Nevertheless, in the
problems with the effective dimensional reduction as in
\cite{GMS1,GMS2,GMS3}
( assumed that true dimension in greater than $0+1$ or $1+1$,
respectively ) the result obtained in the leading order in $1/N$
expansion retains to be correct even after considering next to leading
order, i.e. $1/N$ perturbation theory does not fail. Although the
above statement can be rigorously proved, that is not our aim. Here
we only present a physical reason which could  clearify
this subtle point. Since the dimensional reduction takes place
only in charged channels, it affects only the fermion-antifermion
pairing dynamics. The Nambu-Goldstone (NG) bosons are neutral
and there is no any sign of dimensional reduction for them. So the
argument that NG bosons, as a result of D$\chi$SB, cannot appear
at low dimensions is not decisive now and D$\chi$SB really takes place.

Bellow we shall study the gap equation in the problem. As was indicated
above, the emphasis will be made on the realization of dimensional reduction.
To gain a better understanding of the phenomenon we recall some quite well
known, but very important for our study, facts about D$\chi$SB in NJL
model in an external magnetic field. In the case of Abelian gauge field,
the dependence of dynamical masses in NJL model in different dimensions
on the coupling constant (see, for example, Refs.\cite{GMS1,GMS2}) is given
by:
\xq
m_{dyn}\simeq\frac{|eB|}{2\pi}NG_0, \qquad \mbox{at}\qquad  D=2+1,
\label{eq:d21}
\zq
\xq
m_{dyn}\simeq\frac{|eB|}{\pi}\exp\left(\frac{\Lambda^2}{|eB|}\right)
\exp\left(-\frac{4\pi^2}{|eB|NG_0}\right), \qquad \mbox{at}\qquad  D=3+1.
\label{eq:d31}
\zq
Both cases correspond to a small coupling parameter, $G_0\to 0$. Were the
field not applied to the systems, the D$\chi$SB would be absent at any
coupling less than the critical value. The external magnetic field, as we
can see from (\ref{eq:d21}) and  (\ref{eq:d31}), changes the situation
drastically: the D$\chi$SB in magnetic field appears to take place at any
nonzero attraction in particle--antiparticle channel. Such a situation in
the aforecited
papers was interpreted as a result of the dimensional reduction in fermion
pairing dynamics by two units.  Actually, we can easily check that for
NJL models without  magnetic fields in $0+1$ and $1+1$ dimensions, in
leading order in $1/N$ perturbation theory, the dynamical masses depend on
the coupling in the same way as in (\ref{eq:d21}) and  (\ref{eq:d31}),
respectively. The only difference is connected with the mass
scales in these two approaches. In pure $(0+1)$-- and $(1+1)$--dimensional
models the mass scale is completely determined by the ultraviolet cutoff.
The NJL models reduced by means of magnetic field contain also another
scale related to the magnetic field, thus, it is not surprising that in this
case it interplays with the cutoff and enters into
(\ref{eq:d21}) and  (\ref{eq:d31}).

As for the case of chromomagnetic field, it is known that in $2+1$ dimensions
the dynamical mass depends on the coupling in the following way \cite{Vshiv}:
\xq
m_{dyn}\simeq\sqrt{gH}\exp\left(\frac{6\Lambda}{\sqrt{gH}}\right)
\exp\left(-\frac{2\pi}{G_0N\sqrt{gH}}\right).
\label{eq:d21c}
\zq
After comparing this expression with (\ref{eq:d31}), the idea of
dimensional reduction simply suggests itself. But now the reduction,
if it does really exist, in contrast to the
case of Abelian field, turns out to be only by one unit ($2+1\to 1+1$).

So the main aim of our paper is to prove the hypothesis that a chromomagnetic
field really causes the dimensional reduction in NJL model by one unit.

There is, of course, another possibility. One could assume that a
chromomagnetic field acts in the same way at any dimensions,
{\em i.e.} it always causes appearance of D$\chi$SB even at the lowest
attraction in particle--antiparticle channel. However, we shall see
that it is not so.

To start with let us consider NJL model in an external $SU(3)$ chromomagnetic
field:
\xq
{\cal L} = \bar{\Psi}_k i\gamma^\mu D_\mu\Psi_k+
\frac{G_0}{2} \left[ (\bar{\Psi}_k\Psi_k)^2+(\bar{\Psi}_ki\gamma^5\Psi_k)^2
\right],
\zq
where $D_\mu=\partial_{\mu}-igA_{\mu}^{a}T^{a}$ is the covariant derivative.
In order to make use of the $1/N$ pertubation theory, we use the model with
$N$ flavors, {\em i.e.} fermion field carries an additional index
$k=1,2,\dots,N$. The
summation over the repeating indices is assumed. To be specific,
let us choose
the generators of $SU(3)$ group in the form: $T^a=\lambda^a/2$, where
$\lambda^a$ are the Gell--Mann matrices. Here we consider the problem in
$3+1$ dimensions. As for the Dirac algebra, it will be enough only to know
that a four--dimensional irreducible representation is used. Since it is not
very important for our purposes to investigate the case of the most general
configuration of a chromomagnetic field, we assume that the only nonzero
component of the field is $F^{3}_{12}=H$. To go further, we should note that,
as is known, this choice of field does not fix a kind of vector potential
needed. There can be two nonequivalent classes of vector potentials which
describe different physics, though, at the same time, they produce
identical fields \cite{Brown}. The first one corresponds to the vector
potential producing the field which is equivalent to a constant Abelian
field with a linear potential:
\xq
A_{\mu}^{a}=H\delta_{\mu}^{2}x^1\delta_{3}^{a}.
\zq
The physics of fermions in such an external field is totally equivalent
to the case of an ordinary Abelian magnetic field, considered in \cite{GMS2}.
The second possible choice is so called noncommuting potential \cite{Brown},
which give rise to a completely different physics of fermions. The explicit
form of the second type vector potential is:
\xq
A_{1}^{1}=A_{2}^{2}=\sqrt{\frac{H}{g}}\neq 0,
\label{eq:vecpot}
\zq
and all other components are equal to zero.

Performing the standard trick of introducing an auxiliary
Habbard-Stra\-ta\-no\-vich fields, we come to the equivalent lagrangian:
\xq
{\cal L} = \bar{\Psi}_k\left( i\gamma^\mu D_\mu
-\sigma-i\gamma^5\pi\right)\Psi_k
- \frac{1}{2G_0} \left(\sigma^2+\pi^2\right),
\label{eq:lagr}
\zq
with constrains, $\sigma=- G_0\bar{\Psi}_k\Psi_k$ and
$\pi =-G_0\bar{\Psi}_ki\gamma^5\Psi_k$, which are the consiquences of
Euler--Lagrange equations. The last representation is very
convenient for studying D$\chi$SB, since it provides a direct
method for constructing the effective action. Here, we write
down only the expression for effective potential:
\beq
V&=&\frac{1}{2G_0}(\sigma^{2}+\pi^{2})+\tilde{V},\\
\tilde{V}&=&\frac{iN}{\Omega}TrLn(i\gamma^\mu D_\mu
-\sigma-i\gamma^5\pi),
\label{eq:vtilde}
\eeq
where $\Omega$ is the space-time volume and fields $\sigma$ and $\pi$
are assumed to be constant. To find the explicit expression for the
effective potential we can use the momentum representation (see, for example,
Ref.~\cite{Vshiv}). Since the vector potential (\ref{eq:vecpot}) in the
problem does not depend on the space--time coordinates, the differential
operator in (\ref{eq:vtilde}) becomes diagonal in momentum space.
Performing quite simple calculations with matrices and introducing the
notation for the chiral invariant combination of auxiliary fields,
$\rho^2=\sigma^{2}+\pi^{2}$, we obtain the following result:
\xq
\tilde{V}(\rho)=2iN\int \frac{d^4 k}{(2\pi)^{4}} \ln(k_{0}^{2}-E_{+}^{2}
(\vec{k}))
(k_{0}^{2}-E_{-}^{2} (\vec{k}))(k_{0}^{2}-\vec{k}^{2}-\rho^{2}),
\label{eq:Vpot}
\zq
with
\beq
E_{\pm}^{2}&=&k_{\perp}^{2}+k_{3}^{2}+\rho^{2}+2h^{2} \pm 2h \sqrt
{k_{\perp}^{2}+h^{2}},\\
&&k_{\perp}^{2}=k_{1}^{2}+k_{2}^{2},\qquad  h=\frac{\sqrt{gH}}{2}.
\eeq
The minimum of the effective potential ( as a function of $\rho$ )
determines the value of the order parameter. As for the physical
meaning of the latter, it is obvious already from (\ref{eq:lagr}) that this
value plays a role of the fermion mass which is dynamically
generated due to 4F interaction.  In the relativistic physics
the mass coincides with the half of the gap in the fermion spectrum
( the gap between antiparticle and particle states ), so the equation
for the effective potential minimum is often refered to as the gap
equation. In our problem it looks like:
\xq
\frac{\rho}{G_{0}}+\frac{\partial \tilde{V} (\rho)}{\partial \rho}=0.
\label{eq:gap0}
\zq
After explicit substitution (\ref{eq:Vpot}) into (\ref{eq:gap0})
we obtain:
\xq
\frac{\rho}{G_{0}}=\frac{i\rho N}{4\pi^{4}} \int d^{4} k \left[
\frac{1}{k_{0}^{2}-E_{+}^{2}}+ \frac{1}{k_{0}^{2}-E_{-}^{2}}
+\frac{1}{k_{0}^{2}-\vec{k}^{2}-\rho^{2}}
\right].
\zq
Following the standart way we rewrite the last expression
in  the Euclidean space ($k_{0}\to ik_{4}$):
\beq
\frac{\rho}{G_{0}}&=&\frac{\rho N}{4\pi^4} \int d^{2} k_{\parallel} d^{2}
k_{\perp}
\bigg[ \frac{1}{k_{\parallel}^{2}+k_{\perp}^{2}+\rho^{2}+2h^{2}
+2h\sqrt{k_{\perp}^2+h^2}}\nonumber\\
&&+\frac{1}{k_{\parallel}^{2}+k_{\perp}^{2}+\rho^{2}+2h^{2}
-2h\sqrt{k_{\perp}^2+h^2}}+\frac{1}{k_{\parallel}^{2}+k_{\perp}^{2}+\rho^{2}}
\bigg],
\label{eq:gap1}
\eeq
where $k_{\parallel}^{2}=k_{3}^{2}+k_{4}^{2}$. Since the expression
(\ref{eq:gap1}) is divergent, we need to use some regularization procedure.
As is common in such problems, to avoid the divergences we shall use cutoff
in momentum space. In our problem the Lorentz invariance already is broken
by an external chromomagnetic field and we may choose, in principle,
two different cutoffs for $k_{\parallel}$ and $k_{\perp}$. In order to
avoid unnecessary complication we assume that
$0\leq k_{\parallel}, k_{\perp}\leq\Lambda$.

Performing integration in (\ref{eq:gap1}) and omitting all terms of
order $O(1/\Lambda)$, we can rewrite the gap equation in the
form:
\beq
\Lambda^2\left(\frac{1}{g}-\frac{1}{g_c}\right)&=&-\frac{\rho^2}{2}
\left[\ln\frac{\Lambda^2}{2\rho^2}+1-\frac{1}{3}\ln\frac{\rho^2+4h^2}{\rho^2}
\right]-
\nonumber\\
&&-\frac{h^2}{3}\left[1+\frac{2\rho}{h}\arctan\frac{2h}{\rho}\right],
\label{eq:gap}
\eeq
If there is no real solution to this equation, then $\rho=0$. The
dimensionless coupling constant, $g$, was defined by the expression:
\xq
g=\frac{3}{2\pi^2}\Lambda^2G_0N
\zq
and the "critical" value $g_c$ equals to $1/\ln{2}$ (note that this value
depends on the incorporated cutoff and, as a result, it differs from
the critical value in the problem without external fields where
the commonly used cutoff reads as
$0\leq k_{\parallel}^2+ k_{\perp}^2\leq\Lambda^2$). The physical
meaning of $g_c$ becomes obvious when the external field is switched
off. In this case the gap equation becomes very simple:
\beq
\Lambda^2\left(\frac{1}{g}-\frac{1}{g_c}\right)&=&-\frac{\rho^2}{2}
\left[\ln\frac{\Lambda^2}{2\rho^2}+1\right].
\eeq
Since the right hand side of this expression is always negative, the
nontrivial solution for the gap appears only at $g>g_c$. So,
D$\chi$SB does not exist at weak attraction in fermion--antifermion
channel, $g<g_c$. If an external chromomagnetic field acted as an
ordinary magnetic field, after applying it to the system with $g\to
0$, we would obtain the D$\chi$SB as a result of the effective
dimensional reduction by two units \cite{GMS1,GMS2}. However, as one
can see from (\ref{eq:gap}), it is not the case: the right hand side of
(\ref{eq:gap}) retains  to be negative ($\Lambda$ is assumed to be
large) even after turning the field on. Thus, the D$\chi$SB does not
take place in an external chromomagnetic field in $3+1$ dimensions at
weak attraction. This conclusion, therefore, does not confirm the
idea about the dimensionally insensitive D$\chi$SB mechanism in a
chromomagnetic field. Actually, although the D$\chi$SB exists at any weak
attraction in chromomagnetic field in $2+1$ dimensions, it disappears when
we add one more dimension.

Now we shall turn our attention to the discussion of the effective
dimensional reduction, namely, to its formal mathematical
manifestation.

For better understanding the phenomenon of dimensional reduction,
let us compare the gap equation (\ref{eq:gap1}) with the corresponding
expression
in the case of ordinary magnetic field \cite{GMS2}:
\beq
\frac{\rho}{G_0}&=&\frac{\rho N}{4\pi ^{4}}\int d^{2} k_{\parallel}
d^{2} k_{\perp} \exp\left(-\frac{k_{\perp}^{2}}{|eB|}\right)
\sum_{n=0}^{\infty} \frac{L_{n}^{-1}(2k_{\perp}^{2}/|eB|)}
{k_{\parallel}^{2}+\rho^{2}+2|eB|n}=\nonumber\\
&=&\frac{\rho N|eB|}{4\pi ^{3}}\int d^{2} k_{\parallel}
\sum_{n=0}^{\infty} \frac{2(-1)^n-\delta_{n,0}}
{k_{\parallel}^{2}+\rho^{2}+2|eB|n}.
\label{eq:zero}
\eeq
The last expression explicitly shows that the effective dimensional
reduction really takes place in the system. The only feature in
(\ref{eq:zero}) that spoils similarity with pure two--dimensional model
(without external fields) is the sum over Landau levels. However, there is
one very important case when this sum does not play any crucial role.
Such a case is realized when the attraction in particle--antiparticle
channel is weak, $G_0\to 0$. Then the dynamical mass will be rather small
(in comparison with magnetic field scale), $\rho\to 0$, and the predominant
contribution will come from the lowest Landau level, $n=0$, {\em i.e.}
\beq
\frac{\rho}{G_0}\simeq\frac{\rho N |eB|}{4\pi ^{3}}\int \frac{d^{2}
k_{\parallel}}
{k_{\parallel}^2+\rho^2}=
\frac{\rho N|eB|}{4\pi^{2}}\left[\ln\frac{\Lambda^2}{\rho^2}+
O\left(\frac{\rho^2}{\Lambda^2}\right)\right].
\eeq
It is this the lowest Landau level contribution that determines the
exponential factor in (\ref{eq:d31}) with nonanalytical dependence on the
coupling constant.

{}From the  mathematical point of view the appearance of D$\chi$SB at
$G_0 \to 0$ caused by divergence of the integral in (\ref{eq:zero}) in the
infrared region ($k_{\parallel}\to 0$) when the dynamical mass, $\rho$,
tends to zero. In this $(3+1)$--dimensional model the divergence is very weak
(only logarithmic), and, as a result, the dynamical mass is exponentially
small. To take a  $(2+1)$--dimensional model, we should change the measure
$d^2k_{\parallel}$ for $dk_{\parallel}$ (except for some unimportant change
of the factor in front of the integral). As a result the infrared divergence
becomes stronger: $\sim 1/\rho$, and, in its turn, the dependence of the
dynamical mass on the coupling constant is linear (see Eq.(\ref{eq:d21})).
These results were obtained in \cite{GMS1,GMS2}.

Turning back to the case of chromomagnetic field, we see that the main
contribution into divergence in the gap equation (\ref{eq:gap1}), comes,
if there is any, from the second term in the integrand. Actually, when the
momentum $k_{\perp}$ tends to zero, the denominator of the second term
approximately is equal to $k_{\parallel}^2+\rho^2$, {\em i.e.} exactly to
the same expression that appears in the case of ordinary magnetic field
due to the lowest Landau level. However, there is a very important difference
between these two cases. It is connected with the dependence of the integrand
on the transverse momentum $k_{\perp}$. For obtaining from the gap equation
the expression of effective lower dimension, in the case of a magnetic field,
one needed only to perform the integration over $k_{\perp}$. As for the
chromomagnetic field, such an integration would lead to an expression which
has not got any resembleness with lower dimensional problems. Thus, to obtain
such a representation of the gap equation (\ref{eq:gap1}) that bears a
resembleness to some lower dimensional problem, we shall perform the
integration only over the polar angle of the momentum $k_{\perp}$. Then
changing the variable $k_{\perp}$ for $k=\sqrt{k_{\perp}^2+h^2}-h$, which
varies from zero to $W=\sqrt{\Lambda^2+h^2}-h$, we can rewrite the second
term in (\ref{eq:gap1}) as:
\beq
\frac{\rho}{G_0}\sim\frac{\rho N}{2\pi ^{3}}\int d^{2} k_{\parallel}
\int\limits_{0}^{W}\frac{(k+h)dk}{k_{\parallel}^2+k^2+\rho^2}.
\label{eq:crost}
\eeq
We see from (\ref{eq:crost}) that the dependence of the
integrand on the momentum in the infrared region is the same as in the
$(2+1)$--dimensional NJL model without external fields, {\em i.e.} there
is no infrared divergences in the system at $\rho\to 0$. This is the reason
why the D$\chi$SB is absent for this case. Thus, the idea of the effective
reduction, at least in regard to the D$\chi$SB, in this $(3+1)$--dimensional
model is not very useful, since it does not lead to the infrared divergences.
The situation in $(2+1)$--dimensional model in external chromomagnetic
field is much more interesting. Without repeating all calculations, we only
note that, at the end, we shall come to the expression for the infrared
divergence similar to the right hand side in (\ref{eq:crost}).
But now the measure
$d^{2} k_{\parallel}$ will be replaced by $dk_{3}\equiv -idk_{0}$ and the
constant before the integral will be another. The infrared divergence at
$\rho\to 0$ appears as a result of changing the  measure. Omiting the
convergent term proportional to $k$ (as in (\ref{eq:crost})), we obtain the
following:
\beq
\frac{\rho}{G_0}\sim\frac{\rho Nh}{2\pi^2}\int d k_{3}
\int\limits_{0}^{W}\frac{dk}{k_{3}^2+k^2+\rho^2}\sim
\frac{\rho Nh}{4\pi^2}\int \frac{d^2q}{q^2+\rho^2}.
\label{eq:end}
\eeq
This expression obviously proves our hypothesis about the effective
dimensional reduction by one unit ($2+1\to 1+1$) when an external
chromomagnetic field is
applied to the system. It is just this logarithmic infrared divergence of
the right hand side in (\ref{eq:end}) at $\rho\to 0$ that causes the
appearance of the D$\chi$SB in the system and the exponential nonanalytic
dependence of the dynamical mass on the coupling constant as in
(\ref{eq:d21c}).

In conclusion, we repeat once again the main result of the paper. The
influence of an external chromomagnetic field on the fermion system with the
weak attraction in particle--antiparticle channel (as an example, NJL model
was used) was studed. It was found that such a model considered in $D$
dimensions behaves, in regard to the particle--antiparticle pairing dynamics,
like the model without external fields but in $D-1$ dimensions. Our analysis
obviously shows that the infrared region of the fermion spectrum in an
external field plays a crucial role in appearing of this phenomenon, which we
call the effective dimensional reduction. In contrast to a similar phenomenon
in an external Abelian magnetic field with effective dimensional
reduction by two units ($D\to D-2$), a chromomagnetic field reduces the
number of dimensions only by one unit ($D\to D-1$).

The research was supported in part
(for I.A.Sh.) by the International Soros Science Education Program
(ISSEP) through grant No.PSU052143.

\end{document}